\documentclass[structabstract]{aa}
\usepackage{graphicx}
\usepackage{color}
\usepackage{epsfig}
\usepackage[varg]{txfonts}
\usepackage{natbib}
\usepackage[utf8]{inputenc}
\bibpunct{(}{)}{;}{a}{}{,}
\def\kms {\rm{km~s^{-1}}}

\def\apjl {ApJL}
\def\apjs {ApJS}

\def\aap {A\&A}

\def\Mpc {\rm Mpc}

\def\AA {\r{A}}
\begin{document}

\title{The BL-Lac gamma-ray blazar PKS 1424+240 associated with a group of galaxies at z=0.6010}

\titlerunning{The blazar PKS 1424+240 at $z=0.6010$}

\author{A.C. Rovero \inst{1}, H. Muriel \inst{2,3}, C. Donzelli \inst{2,3} \and A. Pichel \inst{1}}

\institute{Instituto de Astronom\'ia y F\'isica del Espacio 
(IAFE, CONICET-UBA), Av. Inte. G\"uiraldes 2620, C1428ZAA, Ciudad Aut\'onoma de Buenos Aires, Argentina \and
Instituto de Astronom\'ia Te\'orica y Experimental (IATE, CONICET-UNC), Laprida 854, X5000BGR, 
C\'ordoba, Argentina \and Observatorio Astron\'omico, Universidad Nacional de C\'ordoba, 
Laprida 854, X5000BGR, C\'ordoba, Argentina \\
         \email{rovero@iafe.uba.ar}
}

\date{Received ... ; accepted ...}

%\abstract{}{}{}{}{} 
% 5 {} token are mandatory

\abstract
 % context heading (optional) leave it empty if necessary  
   {PKS 1424+240 is a BL-Lac blazar with unknown redshift that was detected at high-energy gamma rays by Fermi-LAT with a hard spectrum. At very high energy  (VHE), it was first detected by VERITAS and later confirmed by MAGIC. Its spectral energy distribution is highly attenuated at VHE gamma rays, which is coherent with distant sources. Several estimations enabled the redshift to be constrained to the range $0.6 < z < 1.3$. 
These results place PKS 1424+240 in the very interesting condition of being probably the most distant blazar that has been detected at VHE. The ambiguity in the redshift is still large enough to prevent precise studies of the extragalatic background light and the intrinsic blazar spectrum.}
%   aims heading (mandatory)
   {Given the difficulty of measuring spectroscopic redshifts for BL-Lac objects directly, we aim to establish a reliable redshift value for this blazar by finding its host group of galaxies.}
%   methods heading (mandatory)
   {Elliptical galaxies are associated with groups, and BL-Lac objects are typically hosted by them, so we decided to search for the host group of the blazar. For this, we performed optical spectroscopic observations of thirty objects in the field of view of PKS 1424+240 using the Gemini Multi-Object Spectrograph. After analysing the data for groups, we evaluated the probability of finding groups of galaxies by chance around the position of PKS 1424+240, using a deep catalogue of groups. We also used photometric data from the SDSS catalogue to analyse the red sequence of the proposed blazar host group}
%   results heading (mandatory)
   {We found a new group of galaxies with eight members at $z = 0.6010 \pm 0.003$, a virial radius of $R_{vir}=1.53~\Mpc,$ and a velocity dispersion of $\sigma_v = 813 \pm 187$ km/s. The photometric study indicates that more members are probably populating this previously uncatalogued group of galaxies. The probability of PKS 1424+240 being a member of this group was found to be $\gtrsim 98\%$.}
%   conclusions heading (optional), leave it empty if necessary 
{The new group of galaxies found at $z =0.6010 \pm 0.003$ is very likely hosting PKS 1424+240.}


\maketitle
%
%________________________________________________________________

\section{Introduction}
\label{sec:intro}
%------------------------------------
   \begin{figure*}
   \centering
   \includegraphics[width=10cm]{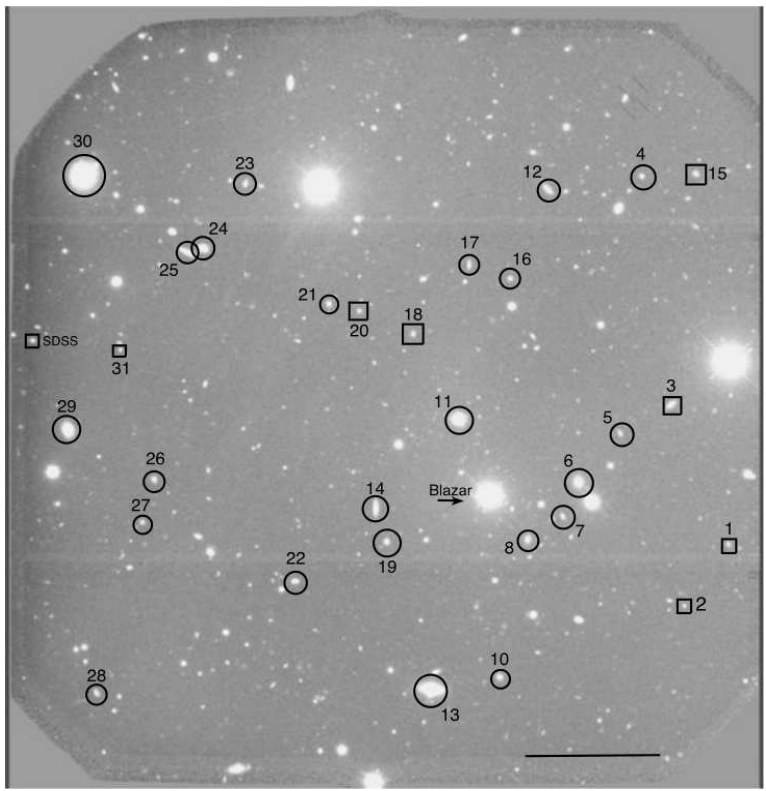}
   \caption{GMOS $r'$-band image of the PKS 1424+424 field. North is at the top and east to the left. Spectroscopically observed object are labelled according to the slit number (see Table \ref{tab:data}). Squares indicate objects that belong to the probable blazar group, while circles show foreground or undetermined redshift objects. The lower right bar indicates 1 arcmin.}
   \label{fig:gemini}
   \end{figure*}
%------------------------------------

Redshift determination of gamma-ray sources is a challenge in extragalactic gamma-ray astronomy, which is mostly populated by BL-Lac type blazars. Blazars are active galactic nuclei (AGN) with strong emissions that are produced at all wavelengths in the relativistic jets, and which are oriented along the line of sight. Blazars are populated by two types of objects: flat-spectrum radio quasars and BL Lacertae (BL-Lac).

Spectroscopic redshifts are difficult to determine for BL-Lacs. The absorption and emission optical lines that are used to determine the redshift of the host galaxy are not strong enough to rise over the non-thermal continuum from the jet (e.g. \citealt{landt02}; \citealt{sbarufatti05}).
The lack of redshift determination for extragalactic gamma-ray sources severely limits the modelling of the source because the very high-energy (VHE; E$>$100 GeV) gamma radiation that reaches the Earth is attenuated by the extragalactic background light (EBL), the effects of which depend on the distance (e.g. \citealt{Reimer:2013}).
The EBL is a diffuse cosmological radiation field that covers the UV to far-IR, which encompasses all the radiative energy releases since recombination, and is dominated by the formation of massive stars.
Gamma rays interact with low-energy ambient EBL photons producing an electron-positron pair, which is the main attenuation effect for extragalactic VHE gamma-ray astronomy (e.g. \citealt{stecker:1992}). Consequently, VHE sources that are detected are relatively close (typically $z <0.6$, except for the cases of lensing effect, e.g. \citealt{sitarek:15}). At these redshifts, the attenuation effect for gamma rays at high-energy (HE; E$>$100 MeV) is negligible. Thus, by extending the spectrum from HE to VHE, a measure of the absorbed gamma-ray radiation may be obtained, which in turn can be used to estimate either the spectral properties of the EBL or the redshift of the source (e.g. \citealt{aharonian06}; \citealt{albert08}; \citealt{finke09}; \citealt{orr11}). Results obtained using this procedure are quite uncertain, owing to the uncertainty of the EBL density and because extending the SED from HE to VHE involves making strong assumptions (e.g. \citealt{dwek13}).

Other methods for estimating redshift  include the analysis of absorption lines caused by intergalactic matter. The red shifted spectra of far-away sources show absorption lines that are caused by hydrogen clouds in the object line of sight (the Ly forest). By analyzing the redshifts of those lines, it is possible to find lower and upper limits to the redshift of the source (e.g. \citealt{Danforth:2010, Danforth:2013, Furniss:2013}). This method requires far-UV spectra, which are not easy to get as they must be obtained from space-borne observatories.

With a totally different approach, in a recent publication we proposed an alternative method to estimate the redshift of BL-Lac blazars in an indirect way. 
Elliptical galaxies are known to be combined to form groups or clusters, and also that BL-Lacs are typically hosted by elliptical galaxies. We therefore
 proposed to analyze spectroscopic observations to find the host group of galaxies that are associated with the blazar (see \citealt{muriel15} for details). This method needs to correctly estimate the probability of detecting a group of galaxies by chance, i.e. of not being the one the blazar belongs to, which strongly depends on the type of observations used for the study.

PKS 1424+240 is a BL-Lac blazar (type HBL) with unknown redshift detected at HE by Fermi-LAT with a hard spectrum \citep{abdo09}. It was first detected at VHE by VERITAS (VER J1427+237) \citep{acciari10}, and later confirmed by MAGIC \citep{aleksic14}. 
Modeling the change from HE to VHE in the spectral index of PKS 1424+240, an upper limit of $z < 0.66$ was found by considering different EBL models \citep{acciari10}. 
Using a statistical approach that correlates the drop in the gamma-ray spectra to real redshift measurements from known sources, \cite{prandini:11_b} estimated a probable redshift for PKS 1424+240 of $z \sim 0.26$, with an upper limit of $z < 0.45 \pm 0.15$, although this result is based on the assumption that all blazars behave the same way. 
\cite{yang:2010} found an upper limit of $z < 1.19$ using only the lowest EBL estimation.
\cite{scully:2014} determined a conservative upper limit of $z < 1.0$ using new EBL data derived from deep galaxy surveys.
A two-component synchrotron self-Compton model was found to well describe the SED of the source if it is located at $z \sim 0.6$ \citep{aleksic14}.
Considering PKS 1424+240 as a source of ultra-high-energy cosmic rays, models were consistent with redshift ranges of $0.6 < z < 0.75$ \citep{yan_a:15}, and $0.6 < z < 1.3$ \citep{yan_b:15}.
Leaving the gamma-ray domain, a photometric upper limit of $z < 1.11$ was reported by \cite{rau12}. Also, a totally independent and more firm redshift lower limit $z \geq 0.6035$ was reported for PKS 1424+240 by \cite{furniss13}. They analyzed the Ly forest from recent UV observations taken with the HST/COS, which covers the range 1135-1795 \AA. All these results place PKS 1424+240 in the very interesting condition of being one of the few most distant blazars detected at VHE, with redshift in a range never populated by other VHE blazars, i.e. $0.6 < z < 1.3$. This ambiguity in the redshift is still large enough to prevent precise studies of the EBL and the intrinsic blazar spectrum.\\

We used Gemini for spectroscopic observation of the blazar PKS 1424+240 and other objects in the field of view. Spectroscopic redshift from the blazar spectrum could not be determined \citep{rovero15}. In this paper, and following the idea outlined in a previous publication \citep{muriel15}, we analyze spectroscopic observations of 30 objects in the environment of the blazar, seeking its host group of galaxies. In Section \ref{sec:Obs}, we describe the observations and data reduction; in Section \ref{sec:Res}, we present the analysis and results, and Section \ref{sec:Con} summarizes the conclusions.

A cosmology was applied with $H_{0} = 70$ $\kms$  $Mpc^{-1}$, ${\Omega }_{m} = 0.25$, and ${\Omega }_{\Lambda} = 0.75$.

%------------------------------------
\section{Observations and data reduction}
\label{sec:Obs}

Spectroscopic observations were carried out with the Gemini North telescope using the Gemini
Multi Object Spectrograph (GMOS). These observations were acquired under the Gemini program
 GN-2015A-Q12 (PI: A.C. Rovero). A multislit mask was created using a 60 s exposure pre-image
 in the $r'$ filter taken on March 15 2015 (Figure \ref{fig:gemini}). This image covers 5 x 5 arcmin$^2$, with a pixel scale
 of 0.146 arcsec. Objects were selected by eye around the field when their integrated
 magnitudes were not fainter than $m_r' = 22.5$ to get spectra with a reasonable
 S/R $> 5$ ratio. We also selected the centre and the position angle of the image 
 to maximize the slit number in the mask. We were able to allocate 31 slits of 1 arcsec
 width and 4 arcsec length. Spectra were taken in queue mode on April 26 2015, under excellent
 seeing conditions (FWHM $\sim 0.7$ arcsec). A total of 5 x 900 s exposures were obtained
 at three different central wavelengths (540 nm, 550 nm, and 560 nm) so as to correct for 
the gaps between CCDs. The B600$\pm$G5323 grating was used with a dispersion of 
$\sim$0.9 $\AA$ per pixel and a resolution of FWHM $\sim 5 \AA$. The spectra cover the 
range $4000-7000 \AA$, but the exact range for each spectrum depends on the slit position 
on the mask. Image and spectra reduction followed the standard procedures, using IRAF and the Gemini IRAF package. Further details 
may be seen in \cite{rovero15}. Similarly, redshifts were measured using the FXCOR 
task under IRAF. Details on this procedure can be found in \cite{muriel15}.

\section{Results}
\label{sec:Res}

\begin{figure}
   \centering
   \includegraphics[width=8.5cm]{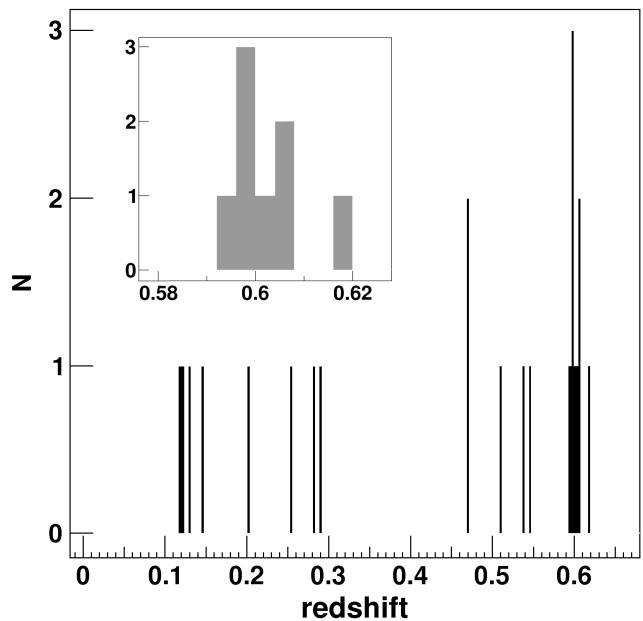}
   \caption{Distribution of redshifts for the galaxies observed in the field of view around PKS 1424+240. Inset: narrower range around the group of seven members surrounding the blazar.}
   \label{fig:z}
   \end{figure}

Results of the spectroscopy performed on the targeted objects are summarised in Table \ref{tab:data}. Figure \ref{fig:z} shows the distribution of redshift measured in this work for the 21 galaxies in the sample. Within our field of view there are other galaxies included in the spectroscopic Sloan Digital Sky Survey SDSS catalog; three of them are the galaxies in slits 8, 13, and 30 (see Table \ref{tab:data}), whose reported redshifts are 0.4688, 0.1212, and 0.1195, respectively, which confirm our measurements. There is a fourth galaxy at RA(J2000.0) 216:48:55.5 and Dec(J2000.0) 23:48:51.7, which is not included in our sample, with a redshift of $z = 0.6047$. We are including this galaxy in our analysis.

\begin{table*}
\caption{Targets with spectroscopic observations used in this paper. Column 1: slit number; Columns 2 and 3: RA and Dec (J2000.0); Column 4: total $r'$ integrated magnitude; Column 5: redshifts estimated in this work.}
\center
\begin{tabular}{rcclll}
\hline \hline
Slit  & RA (J2000.0) & Dec (J2000.0) & $m_r'$ & Redshift  & Comments \\
%(1)   &   (2)     &  (3)         &  (4)  & (5)                         & (6) \\
\hline 

1  &  14:26:52.5 &  23:47:47.7  &  22.2  & z = 0.5984 $\pm$ 0.0001 & \\
2  &  14:26:53.8 &  23:47:18.5  &  22.6  & z = 0.5999 $\pm$ 0.0006 & \\
3  &  14:26:54.8 &  23:48:48.2  &  20.3  & z = 0.5970 $\pm$ 0.0001 & \\
4  &  14:26:56.4 &  23:50:29.5  &  21.5  & z = 0.2547 $\pm$ 0.0009 & \\
5  &  14:26:56.4 &  23:48:33.3  &  21.8  & ~---  & \\
6  &  14:26:57.6 &  23:48:09.7  &  19.1  & z = 0.2003 $\pm$ 0.0007 & \\
7  &  14:26:58.0 &  23:47:53.8  &  21.4  & z = 0.6163 $\pm$ 0.0001 & \\
8  &  14:26:59.0 &  23:47:42.0  &  20.5  & z = 0.4688 $\pm$ 0.0005 & \\
9  &  14:27:00.4 &  23:48:00.4  &  ~---  & ~---  & PKS 1424+240 \\
10 &  14:26:59.6 &  23:46:39.1  &  21.0  & z = 0.4692 $\pm$ 0.0005 & \\
11 &  14:27:01.7 &  23:48:33.1  &  18.8  & ~---  & \\
12 &  14:26:59.4 &  23:50:20.0  &  20.6  & ~---  & \\
13 &  14:27:01.8 &  23:46:31.1  &  17.5  & z = 0.1218 $\pm$ 0.0002 & \\
14 &  14:27:04.1 &  23:47:49.8  &  20.1  & ~---  & \\
15 &  14:26:54.6 &  23:50:32.6  &  21.5  & z = 0.6078 $\pm$ 0.0003 & \\
16 &  14:27:00.4 &  23:49:38.4  &  21.6  & ~---  & \\
17 &  14:27:01.7 &  23:49:43.0  &  21.3  & ~---  & \\
18 &  14:27:03.4 &  23:49:10.0  &  21.6  & z = 0.6018 $\pm$ 0.0007 & \\
19 &  14:27:03.6 &  23:47:35.5  &  20.4  & z = 0.1283 $\pm$ 0.0001 & \\
20 &  14:27:05.2 &  23:49:17.9  &  21.6  & z = 0.6042 $\pm$ 0.0003 & \\
21 &  14:27:06.2 &  23:49:19.9  &  21.4  & z = 0.5470 $\pm$ 0.0003 & \\
22 &  14:27:06.5 &  23:47:14.2  &  21.0  & z = 0.5104 $\pm$ 0.0005 & \\
23 &  14:27:09.2 &  23:50:10.4  &  20.2  & ~---  & \\
24 &  14:27:10.4 &  23:49:40.1  &  19.8  & z = 0.2800 $\pm$ 0.0003 & \\
25 &  14:27:10.9 &  23:49:37.5  &  19.8  & ~---  & \\
26 &  14:27:11.4 &  23:47:54.1  &  21.7  & z = 0.2883 $\pm$ 0.0003 & \\
27 &  14:27:11.6 &  23:47:34.5  &  21.9  & z = 0.5360 $\pm$ 0.0006 & \\
28 &  14:27:12.6 &  23:46:16.1  &  22.0  & ~---  & \\
29 &  14:27:14.3 &  23:48:13.1  &  18.4  & z = 0.1478 $\pm$ 0.0002 & \\
30 &  14:27:14.5 &  23:50:07.6  &  17.2  & z = 0.1178 $\pm$ 0.0002 & \\
31 &  14:27:12.8 &  23:48:50.9  &  22.3  & z = 0.5949 $\pm$ 0.0008 & \\
\hline

\end{tabular}
\label{tab:data}
\end{table*}

%------------------------------------

%------------------------------------

From Figure \ref{fig:z}, it is apparent that there are three possible galaxy systems at $z \sim 0.60$, $z \sim 0.47$, and $z \sim 0.12$, respectively. The first is a group of eight galaxies, seven from our observations (at slits 1, 2, 3, 15, 18, 20, and 31) and the one added from the SDSS spectroscopic catalog.
For this group, using the $Gapper$ estimator by \citet{Beers:1990}, we computed both $z_{mean}$ (mean redshift) and $\sigma_v$ (velocity dispersion) to be $0.6010$ and $813 \pm 187$ km/s, respectively. The virial radius was estimated as $R_{vir}=1.53$ $\Mpc$ (using \citealt{Nurmi:2013}).
We note that both the virial radius and the velocity dispersion are somewhat high, which suggests that this is probably an intermediate-mass cluster rather than a group of galaxies. This led us to explore the red sequence to find evidence of a richer system (see Section \ref{sec:red}).
The pair at $z \sim 0.47$ is formed by galaxies at slits 8 and 10, with a projected distance of 0.38 $\Mpc$ and a difference in velocity of 81 km/s.
The last possible structure is formed by galaxies at slits 13, 19, and 30. The differences in speed between any two of them are 1733 km/s, 1071 km/s, and 2804 km/s, which are too high to be considered as a physical association. 
We note that the limitation on the number of objects to be observed simultaneously with the GMOS implies that the number of members reported here should be considered as a lower limit.

\subsection{Membership of PKS 1424+240}
\label{sec:member}

Within the restricted range of redshifts for PKS 1424+240, as discussed in previous sections, the new group of eight members at $z=0.6010$ found in this work would be the natural host for the blazar. It could be argued that our observations are not deep enough to rule out larger redshifts for PKS 1424+240. However, we show in this section that the probability of finding a group of eight or more members ($P_8$) in an observation like the one reported here 
by chance  is sufficiently low to consider this group the host of the blazar. To evaluate this probability we use the zCOSMOS 20k group catalog (\citealt{knobel12}),
an optical group catalogue in the redshift range $0.1 \lesssim z \lesssim 1.0$, with 1~498 identified groups, 192 of which contain more than five observed members, in $\sim$ 1.7 deg$^2$ of the COSMOS field \citep{Scoville:2007}. We use the catalogue that is restricted to the most complete inner field in its field of view (149.58 deg < RA < 150.66 deg; 1.76 deg < Dec < 2.68 deg), which we refer to as the 20k catalog.
The idea of the procedure to evaluate $P_8$ is, within the field of the 20k catalogue, to select random positions and see the coincidence with groups with eight or more members in the catalogue, considering the groups as circle targets with virial radius as defined by \cite{knobel12}.

Before computing the probability mentioned above, there are considerations to take into account: (i) the 20k catalogue is deeper than the observation reported in this work, so, if we use this catalogue as it is, $P_8$ would be overestimated; (ii) the sample we  took from our field of view was limited by the number of slits we could accommodate (30, excluding the blazar itself), which was completed with the inclusion of one galaxy from the SDSS spectroscopic catalog. We call these 31 objects the observed sample. A reduced sampling of the field introduces fluctuations, i.e. the number of members of the groups found in this work would vary if we chose a different sample, which must be included in the probability estimation.  

To account for the first of these limitations, we  constructed a new catalogue (the so-called pruned catalogue) by selecting galaxies from the 20k catalogue that  have approximately the same optical magnitude distribution as our observed sample. 
By limiting the magnitude of the objects to what we  observed, we  restrict the catalogue to members with redshifts within the depth of our observations. 
A desired second effect is that some galaxies in the 20k catalogue would be too faint for our observations and so the members of their groups would be reduced accordingly in the pruned catalogue.

\begin{figure}
   \centering
   \includegraphics[width=8.5cm]{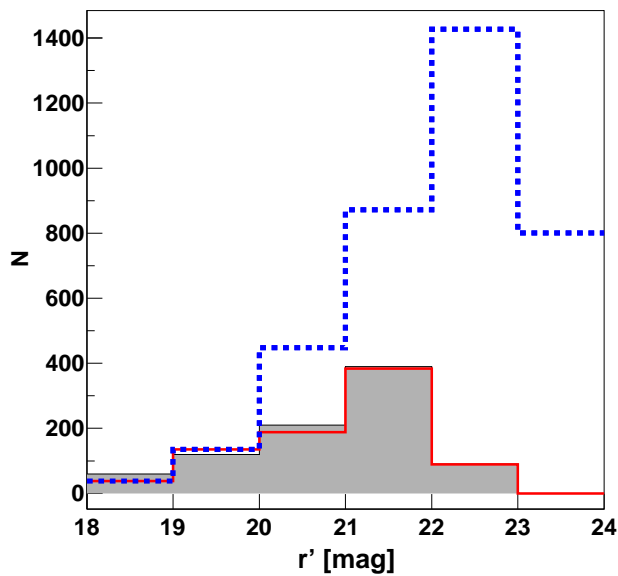}
   \caption{Optical luminosity distribution of galaxies. Blue histogram: galaxies in the 20k catalogue. Black shadow histogram: galaxies observed in this work plus one from the SDSS (the ``observed sample''). The curve was adjusted to match  the blue histogram approximately at high luminosities. Red histogram: one of the pruned catalogues (see text).}
   \label{fig:podado}
   \end{figure}

In Figure \ref{fig:podado}, we show the distributions involved in this process. The blue histogram is the luminosity distribution of galaxies in the 20k catalogue.  
The grey shadow is the luminosity distribution of the observed sample; because we are interested only in the shape of this distribution, the histogram in Figure \ref{fig:podado} has been properly scaled to average the content of the bins at the highest luminosities of the blue histogram (at 18-20 mag).
Then, the construction of the pruned catalogue is made for each luminosity bin (as defined in Figure \ref{fig:podado}) by keeping the ratio between the observed sample to the 20k catalogue; e.g. for the bin 21-22 mag this ratio is $\sim \! \! 0.44$, so we randomly select 44\% of the galaxies from the 20k catalog with luminosities in the range 21-22 mag to populate the corresponding bin of the pruned catalogue.

To account for consideration (ii) above, we constructed 100 pruned catalogues by using different seeds for the random function. This is a natural way to emulate the target selection process in the field of view; a given luminosity bin would be populated by a different set of galaxies for different pruned catalogues. As a consequence, the groups defined in one pruned catalogue are not necessarily the same as in any other; the groups and their number of members are reanalysed for each pruned catalogue. Depending on the richness and seed, the pruned catalogue could have   50\% less groups than in the original catalogue. An example of magnitude distribution of a pruned catalogue is shown in Figure \ref{fig:podado} as a red histogram.

Finally, to estimate the probability of finding a group by chance, we used a Monte Carlo procedure to select random positions within the field of view of the catalogue and, around each of these positions, we searched the pruned catalogues for coincidences with groups of galaxies having a given number of members or more.
Following this procedure we found that the probability of finding by chance
a group of eight or more members in our observations is $P_8 = (7.3 \pm 2.3$)\%, and the error is the standard deviation from 100 values, which corresponds to all pruned catalogues. Considering that the number of members of the new group is a lower limit (see Section \ref{sec:Res}), this probability is an upper limit, i.e. $P_8 < 7.3\%$.

On the other hand, as discussed in our previous work \citep{muriel15}, BL-Lac objects are usually members of galaxy groups. Although the frequency of this membership is difficult to estimate, a value of $\lesssim 0.3$ for the probability of the blazar being isolated would be sensible (e.g. \citealt{Wurtz:1997}, \citealt{Pesce:1995}).
Since this is independent of the probability that is computed above of finding a group by chance, then the probability of PKS 1424+240 not being associated with the group of eight members found in this paper would be the joint probability of having both, i.e. $<0.3 \times 7.3\% \lesssim 2\%$. 
This implies that the possible membership of PKS 1424+240 in the group found at $z=0.6010$ has an occurrence probability of $\gtrsim 98\%$. Moreover, if the most reliable redshift lower limit determined by \cite{furniss13} is taken into account, i.e. $z > 0.6$, the probability for our observations detecting a group with eight or more members {by chance} is practically zero, which would increase the probability of PKS 1424+240 being a member of the newly identified group to nearly 100\%.

%------------------------------------
%------------------------------------

\subsection{Red sequence of galaxies}
\label{sec:red}

\begin{figure}
   \centering
   \includegraphics[width=8.5cm]{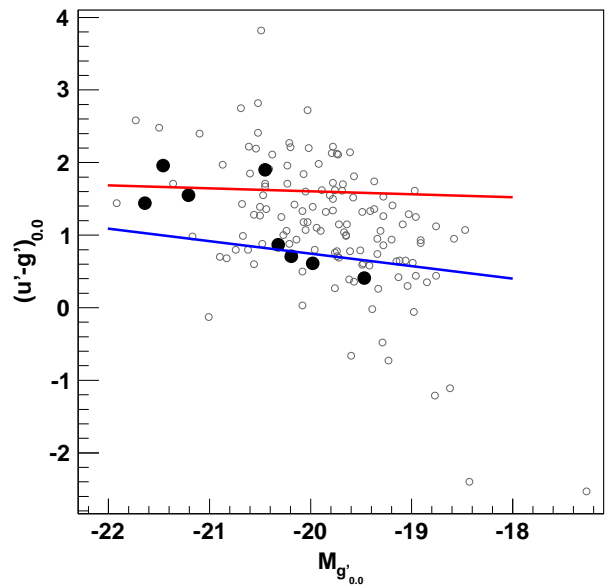}
   \caption{Rest-frame colour-magnitude relation for galaxies in the field of view of our GMOS image, taken from the SDSS photometric catalog. Small circles correspond to galaxies with $z_{phot} > 0.4$, assuming they are at the same redshift as the group ($z \sim 0.6$). Filled circles are the members of the new group found in this work at $z = 0.6010$. The red and blue lines correspond to the red and blue sequence, respectively.}
   \label{fig:red}
   \end{figure}

Many galaxies in massive groups or clusters populate the well-known red sequence (RS; \citealt{visvanathan:77}) in the color-magnitude diagram. A well-defined RS suggests that a system of galaxies is real and evolved. As it was pointed out in Section \ref{sec:Res}, the field around PK 1424+240 overlaps with the SDSS. In addition to the spectroscopic galaxies, there are 210 photometric galaxies in the SDSS field around PK 1424+240, including all our spectroscopic targets. Based on the $u$ and $g$ magnitudes of the SDSS, in Figure \ref{fig:red} we plot the rest-frame color-magnitude relation for the eight spectroscopic members of the group of galaxies at $z \sim 0.6$ (filled circles). The magnitude of galaxies are k-corrected using \cite{blanton:03}. The red and blue lines correspond to the RS and blue sequence (BS) respectively according to \cite{blanton:06} for DEEP2 data \citep{davis:03}. In Figure \ref{fig:red}, we can see that four members fall into the RS region and four into the BS region, suggesting that the group of galaxies at $z \sim 0.6$ is quite evolved.

Using the photometric redshifts ($z_{phot}$) provided by the Twelfth Data Release of the SDSS \citep{alam:15} we selected all the galaxies with $z_{phot} > 0.4$ that fall in the same field of our GMOS image. Small circles in Figure \ref{fig:red} correspond to these galaxies assuming they are at the same redshift as the group ($z \sim 0.6$). We count at least 10 galaxies that are brighter than our faintest spectroscopic RS member falling in the RS region of the group. These galaxies are potential members of the group at $z \sim 0.6$. If this is the case, a larger number of members could help to explain the high values of $R_{vir}$ and $\sigma_v$ reported in Section \ref{sec:Res}. Consequently, the eight spectroscopic galaxies in the group reported in this paper should be taken as a lower limit of its number of members. The expected higher richness, combined with the presence of an evolved population of galaxies, suggests that the 7.3\% probability of finding this group by chance (see Section \ref{sec:member}) should be taken as an upper limit.

%------------------------------------
%------------------------------------

\section{Discussion and conclusions}
\label{sec:Con}

Apart from lensed blazars, PKS 1424+240 is estimated to be the farthest blazar detected at VHE gamma rays. The lack of a firm redshift determination encouraged us to observe this blazar, applying the procedure we used previously for a similar source, i.e. studying its environment to seek its host group of galaxies.
Spectroscopic observations of 30 objects around PKS 1424+240 were obtained using the Gemini Multi-Object Spectrograph. Twenty-one of those were identified as galaxies and their redshift measured spectroscopically in the range $0.11 \lesssim  z \lesssim 0.62$. Seven of these galaxies, plus one added from the SDSS spectroscopic catalog in our field of view, were identified as a galaxy group at $z=0.6010$, with virial radius $R_{vir}=1.53~\Mpc$ and velocity dispersion $\sigma_v = 813 \pm 187$ km/s (which adds a redshift uncertainty of $0.003$). Performing a Monte Carlo analysis on a larger and deeper catalogue of galaxy groups, and considering the limitations of our observations, we estimated the probability of finding a group of eight or more members by chance to be $<7.3\%$. If we add to this estimation the probability of the host galaxy of a blazar being isolated ($\sim 0.3$),
then there is a $\gtrsim 98\%$ probability that the new group of galaxies found at $z =0.6010 \pm 0.003$ is hosting the blazar PKS 1424+240.

The literature contains many attempts to constrain the range of redshift for PKS 1424+240, mostly related to the attenuation suffered by VHE gamma rays when traversing the EBL photon field, a method that presents great uncertainties. Using a much more reliable technique, \cite{furniss13} determined a “strict redshift lower limit of $z \geq 0.6035$ for PKS 1424+240, set by the detection of $Ly_{\beta}$ and $Ly_{\gamma}$ lines from intervening hydrogen clouds''. Their result is remarkably coherent with that found in this paper. Indeed, if we adopt their lower limit as the true value for the redshift of PKS 1424+240, this blazar would be naturally part of the group of galaxies found in this paper. It also seems to be coherent with the fact that \cite{furniss13} have not found HI absorbers between $z = 0.6035$ and the red edge of the detector in $Ly_{\beta}$, i.e. $z = 0.75$, at $\sim 2 \sigma$ confidence level. Moreover, if this strict lower limit is considered, our estimation of the probability of finding a group of eight or more members by chance using our observations would be reduced to nearly zero. This result is reinforced by our photometric study; we concluded that more members should populate this group, which in turn reduces the probability of finding it by chance. 

In summary, we propose that the group of eight or more members found near PKS 1424+240 at $z = 0.6010 \pm 0.003$ is the group that is hosting this blazar, and that very likely the true redshift of PKS 1424+240 is $z = 0.6035$.

%-------------------------------
\begin{acknowledgements}
This work is based on observations obtained at the Gemini Observatory, which is operated by the 
Association of Universities for Research in Astronomy, Inc., under a cooperative agreement with 
the NSF on behalf of the Gemini partnership: the National Science Foundation (United States), 
the National Research Council (Canada), CONICYT (Chile), the Australian Research Council 
(Australia), Minist\'{e}rio da Ci\^{e}ncia, Tecnologia e Inova\c{c}\~{a}o (Brazil) and Ministerio 
de Ciencia, Tecnolog\'{i}a e Innovaci\'{o}n Productiva (Argentina). This work has been partially supported with grants from the Consejo Nacional de Investigaciones 
Cient\'{i}ficas y T\'{e}cnicas de la Rep\'{u}blica Argentina (CONICET) and Secretar\'{i}a de 
Ciencia y Tecnolog\'{i}a de la Universidad de C\'ordoba. We thank H. Julián Mart\'{i}nez for providing the
k-correction of the SDSS magnitudes. The authors are all members of the 
Carrera del Investigador Cient\'ifico of CONICET, Argentina. We thank the anonymous referee for a 
careful reading and useful questions and suggestions that improved the presentation of this paper.
\end{acknowledgements}

%-------------------------------------------------------------------

\bibliographystyle{aa} %*.bst
\bibliography{pks1424} %*.bib

\end{document}